\documentclass{pasj01}
\draft 
\Received{$\langle$reception date$\rangle$}
\Accepted{$\langle$acception date$\rangle$}
\Published{$\langle$publication date$\rangle$}

\usepackage[switch,mathlines]{lineno}
\usepackage{color}
\usepackage{ulem}

\begin{document}

\title{Near-Ultraviolet Radiation toward Molecular Cloud N4 in W 50/SS 433: Evidence for Direct Interaction of the Jet with Molecular Cloud}

\author{Hiroaki \textsc{Yamamoto},$^{*}$ Tatsumi \textsc{Ishikawa}, and Tsutomu, T. \textsc{Takeuchi}}
\affil{Graduate School of Science, Nagoya University, Furo-cho, Chikusa-ku, Nagoya, Aichi 464-8602, Japan}

\email{hiro@a.phys.nagoya-u.ac.jp}


\maketitle

\begin{abstract}

We compared the molecular clouds in the western part of SS 433 with near-ultraviolet radiation data obtained from GALEX. 
Near-ultraviolet radiation is prominently confirmed toward only N4, while no near-ultraviolet radiation is detected toward N1, N2, and N3.
The radiative region of near-ultraviolet radiation is nearly the same as the CO-emitting region in N4, and does not extend beyond the jet seen in X-ray radiation. 
Near-ultraviolet radiation cannot be explained solely by broadband continuous radiation and may originate from line emissions. 
The intensity of near-ultraviolet radiation exhibits an anti-correlation with that of $^{13}$CO($J$=3--2) emission. 
This anti-correlation, along with strong far-infrared radiation in the region with weaker near-ultraviolet radiation intensity compared to its surroundings, suggests that near-ultraviolet radiation originates from behind the molecular cloud, heating up the interstellar dust in N4. 
Subsequently, the dust in N4 reradiates in the far-infrared band. 
In the same region, a high peak $T_{\rm MB}$ ratio of $^{12}$CO($J$=3--2)/$^{12}$CO($J$=1--0) of $\sim$0.9, and a high kinetic temperature of $T_{\rm k}$ $\sim$56 K in the molecular cloud indicate that CO molecules are highly excited, and the molecular cloud is heated through photoelectric heating. 
This heating results from electrons released due to the photoelectric effect caused by the phenomenon where interstellar dust absorbs near-ultraviolet radiation. 
In terms of the timescale of near-ultraviolet radiation originating from line emissions, near-ultraviolet radiation towards N4 cannot be explained by the shock of the blast wave from a supernova that created W 50.
These findings also suggest that N4 directly interacts with the jet from SS 433. 
As a result of this direct interaction, near-ultraviolet radiation is emitted from an interacting layer between the jet and N4.

\end{abstract}

\section{Introduction}

The microquasar SS 433 is well known as an X-ray binary at nearly the center of the supernova remnant W 50. 
SS 433 emanates a powerful jet with a velocity of $\sim$0.26$c$ \citep{mar84}. 
Radio continuum radiation at 1.4 GHz has been identified by \citet{dub98}, and the shell of this radiation is elongated in the east--west direction due to bursts from the powerful jet.

The distance to SS 433 has been estimated in several ways (e.g., \cite{ver93}; \cite{dub98}; \cite{deh98}; \cite{blu04}; \cite{loc07}; \cite{gai18}; \cite{gai21}).
The most reliable estimation of the distance is based on a deep-integration radio image of SS 433 on the arcsecond scale, considering the speed and precession period of the jet (\cite{blu04}; \cite{loc07}).
The estimation places the distance at $\sim$5.5 kpc. 

X-ray observations have revealed the spatial distribution and properties of the jet emanating from SS 433 (e.g., \cite{saf97}; \cite{kot98}; \cite{mol05}; \cite{bri07}). 
The X-ray jet extends by $\sim$0.7 degrees to west and $\sim$1 degree to east, corresponding to $\sim$70 pc and $\sim$100 pc at a distance of 5.5 kpc, respectively. 
High energy gamma-ray radiation in the range of a few hundred GeV to 100 TeV has been detected by Fermi-LAT, HAWC (High Altitude Water Cherenkov Observatory), and more recently LHAASO (Large High Altitude AirShower Observatory)  toward the W 50/SS 433 system (e.g., \cite{abe18}; \cite{fan20}; \cite{cao23}).  
The origin of such GeV-TeV gamma-ray radiation is considered to be inverse Compton scattering.

\citet{yam22} revealed an association of mid-far infrared radiation with molecular cloud N4 (see next paragraph for details of N4), indicating that possibly the interstellar dust in N4 heats up by the interaction of the jet from SS 433. 

CO observations toward the W 50/SS 433 system have been conducted by several authors (e.g., \cite{hua83}; \cite{yam08}; \cite{su018}; \cite{liu20}; \cite{yam22}; \cite{sak23}).
\citet{yam08} identified the molecular clouds associated with the W 50/SS 433 system. 
Four molecular clouds exist in the western part of SS 433, and are lebeled N1--N4.
\citet{yam22} revealed the detailed physical properties of N4, which is located nearest to SS 433 and along same line of sight as X-ray jet. 
The kinetic temperature, $T_{\rm k}$, of the molecular cloud is estimated to be $\sim$56 K at maximum in the western part of N4 from $^{12}$CO($J$=1--0, 3--2) and $^{13}$CO($J$=3--2).
Several observational pieces of evidence indicate that N4 is possibly interacting with the jet from SS 433.

Toward N4, GeV-TeV gamma-ray, X-ray, mid-far infrared, and radio radiation are detected. 
In this paper, we compare the N4 with ultraviolet radiation, which has never been confirmed in the W 50/SS 433 system, and reveal the detailed nature of N4.

\section{Observations}

\subsection{Nobeyama 45m telescope}
Observations were conducted on the molecular clouds detected by NANTEN observations conducted by \citet{yam08} using the Nobeyama 45 m (NRO45m) telescope. 
The front-end system utilized for the observations of the $^{12}$CO($J$=1--0) emission, with a rest frequency of 115.27120 GHz, was BEARS (25-BEam Array Receiver System) \citep{sun00}.
The beam size of the telescope at 115 GHz was $\sim$15$\arcsec$. 
The backend used was Auto Correlators, with 32 MHz bandwidth mode employed for 25 beams \citep{sor00}. 
The velocity resolution was $\sim$0.1 km s$^{-1}$. 
The observations were conducted in the On-The-Fly mapping mode, involving scans along directions of the Galactic longitude and latitude (\cite{man00}; \cite{sa08a}). 
Orthogonal two scans were merged using the basket weaving method \citep{eme88}. 
The observing grid for the OTF mapping mode was 5$\arcsec$, and the grid size of output data matched that of the JCMT archival data at 7$\arcsec$.2.
The data was smoothed to $\sim$20$\arcsec$ to reduce the r.m.s. noise levels. 
The final r.m.s. noise level of the data was $\sim$1.4 K ch$^{-1}$.  
In this paper, we exclusively utilize the dataset pertaining to the N4 cloud in SS 433 as presented in \citet{yam22}.

\subsection{JCMT archival data}
We utilized archival datasets obtained with the JCMT15m telescope. 
The data used in this study corresponds to Proposal ID M18BP027.
In our analysis, we exclusively employed datasets of $^{12}$CO, and $^{13}$CO($J$=3--2) emissions. 
The frequency resolution of the data was 61 kHz, corresponding to the velocity resolution of $\sim$0.05 km s$^{-1}$. For this paper, we further smoothed the velocity resolution to 0.1 km s$^{-1}$. 
The spatial grid of the data was 7\arcsec.2 for the beam size of $\sim$15\arcsec.
The data was smoothed to 20$\arcsec$ to facilinate comparison with $^{12}$CO($J$=1--0).
The main beam efficiency, $\eta_{\rm MB}$ of 0.72 was adopted to convert the intensity scale $T_{\rm a}$ in the archival data to the main beam temperature, $T_{\rm MB}$, scale \citep{buc09}. 
The final r.m.s. noise level for the data of $^{12}$CO, and $^{13}$CO($J$=3--2) was $\sim$1.6, and $\sim$1.9 K ch$^{-1}$, respectively.

\subsection{GALEX archival data}
Image data at UV bands are provided by GALEX (Galaxy Evolution Explorer) and made public. 
GALEX is a UV astronomical satellite launched in 2003 by NASA \citep{mor07}. 
GALEX has far-UV (FUV) and near-UV (NUV) bands and is performing an all-sky survey as well as deep surveys in some sky areas. 
In this analysis, archival datasets from All-Sky Imaging Survey (AIS) GR6/GR7 (e.g., \cite{bia14}) are utilized.  
However, in the W 50/SS 433 region, only the NUV detector was operational, as the FUV detecter did not function initially.
The system angular resolution in the NUV band is 5$\arcsec$.6 at full width at half maximum.
When comparing the NUV datasets with CO datasets, the data was smoothed to 20$\arcsec$. 
The sensitivity of the datasets is $m_{\rm AB}$ $\sim$21 magnitude before smoothing.

\section{Availability of GALEX Data}
Figure 1 shows datasets of GALEX in the field where the present analyzed region is included. 
Figure 1(a) represents an artifact flag map. 
The details are described in the caption of the figure.
In Figure 1(a), four distinct linear structures are seen, traversing the center of the field: two extending from the upper right to lower left and two from the upper left to lower right in yellow.
These are identified as number 17, indicating 'edge + brteddge' artifacts. 
Additionally, a prominent linear structure in green, stretching from the upper-right to lower-left across the center of the field in Figure 1(a), is marked as number 136, indicating 'varmask + varpix'.
These artifact structures are expected to be removed in Figure 1(b).
However, the linear structure around ($\alpha$, $\delta$)(J2000) $\sim$ (\timeform{19h9m56s}, \timeform{5D24'0''}) in Figure 1(b) can not be conclusively identified as either real radiation or a residual error of the artifact. 
This structure lies outside the radio shell of the W50/SS433 system (see Figure 2) and beyond the scope of this paper.
There is an overlap of artifact and radiation near ($\alpha$, $\delta$)(J2000) $\sim$ (\timeform{19h10m20s}, \timeform{5D7'16''}) in the northeastern part of the NUV radiation area. 
The extent of accurate artifact removal is uncertain. However, for the purposes of this study, the NUV radiation around ($\alpha$, $\delta$)(J2000) $\sim$ (\timeform{19h10m20s}, \timeform{5D7'16''}) in Figure 1(b) is considered authentic.

\section{Results}

Figure 2 shows three composite color image, which consists of red: NUV radiation, blue: 0.1--2.4 keV X-ray, and green: 1.4 GHz radio continuum superposed on an integrated intensity map of $^{12}$CO($J$=1--0) emission. 
NUV radiation is prominently confirmed toward N4, while no NUV radiation is observed toward N1, and N2 which are situated at the leading edge of the radio continuum shell. 
\citet{liu20} mentioned that N3 is situated in the radio continuum shell, and at the leading edge of the X-ray jet.
However, no NUV radiation is also seen toward N3.
As mentioned in \citet{yam22}, N4 is situated within the radio continuum shell and along the same line of sight as the X-ray jet emanating from SS 433. 
The distribution of near-ultraviolet radiation in N4 is also confirmed to the same line of sight as the X-ray jet as well, and does not extend beyond that of X-ray radiation.

Notably, strong and extended NUV radiation is concentrated around ($\alpha$, $\delta$)(J2000) $\sim$ (\timeform{19h11m5s}, \timeform{5D16'30''}). 
This extended source of NUV radiation could originate from HD 179124, characterized as a B9V spectral type star, positioned at an estimated distance of $\sim$380 pc from the Sun as inferred from its parallax.
Therefore, this radiation source is unrelated to the W 50/SS 433 system.
While many small, point-like sources are also confirmed toward the W 50/SS 433 system, they are not associated with the W 50/SS 433 system.

The upper-left image in Figure 3 provides an intensity distribution of NUV radiation in N4 superposed on the distribution of peak $T_{\rm MB}$ of $^{12}$CO($J$=1--0) emission, obtained through Gaussian fitting and depicted in contours.
The bright NUV radiating region is split into two parts, west and east. 
The area between these two bright NUV regions corresponds to a local peak in $^{12}$CO($J$=1--0) emission, which we will discuss in more detail in the next section.
The intensity distribution of the region with bright NUV radiation exhibits a patchy nature, with intensity fluctuations less than a factor of 2.
The spectra shown in Figure 3 represent the $^{12}$CO($J$=1--0) spectra for each region, labeled with characters in the upper-left image. 
Additional details regarding these spectra can be found in the caption of Figure 3.
The detection of weak CO emission around the molecular cloud, as defined in \citet{yam22}, suggests that the extent of the molecular cloud is more extensive than what was previously reported in \citet{yam22}, and it closely matches the extent of NUV radiation.
Consequently, this implies a connection between NUV radiation and the molecular cloud. 
Further details regarding the relationship between NUV radiation and the molecular cloud will be discussed in the next section.

\section{Discussion and Summary}
First, we discuss the properties of NUV radiation. 
The radio continuum radiation at 1.4 GHz toward N4 is weak and does not exhibit any morphology related to the shape of the N4 cloud. 
\citet{yam22} have identified mid/far-infrared radiation toward N4, tracing thermal dust radiation with temperature below 100 K, and PAH emission.  
This thermal dust radiation has been detected only in the infrared to radio wavelength, indicating that infrared radiation presented in \citet{yam22} is not related to current NUV radiation.
If radiation from much hotter sources, such as stars with temperatures of thousands of K or more, were present, extended radiation would also be detectable in the near-infrared and visible wavelengths.
However, such extended radiation in the near-infrared and visible wavelengths has not been detected in 2MASS and optical datasets. 
\citet{che02} reported that the broadband spectrum at radio, optical, and UV wavelengths at knot D in 3C279 originates from synchrotron radiation. 
However, in the present case, no counterpart of NUV radiation at other wavelengths is detected, indicating that the origin of NUV radiation is not synchrotron radiation. 
These results suggest that NUV radiation in N4 is not a part of broadband continuous radiation. 
Another possibility for the NUV radiation is emission lines. 
Numerical simulation in \citet{leh20} and \citet{leh22} suggests the presence of UV radiation in the shocked layer.
In the GALEX NUV band, ranging from 1750 $\rm{\AA}$ to 2800 $\rm{\AA}$, strong line emissions such as C$_{\rm III}$] 1909 $\rm{\AA}$ and Mg$_{\rm II}$ 2800 $\rm{\AA}$ are observed.
While these specific line emissions in NUV bands are not included in \citet{leh20} and \citet{leh22}, emissions such as  Ly$\alpha$, Ly$\beta$, and H$\alpha$ emissions are reported. 
Simulations show H$\alpha$ emission in the shocked layer, but towards N4, H$\alpha$ emission has not been detected (details of H$\alpha$ emission are described in the latter half of this section).
C$_{\rm III}$] 1909 $\rm{\AA}$ and Mg$_{\rm II}$ 2800 $\rm{\AA}$ are often detected in distant galaxies and/or QSOs (e.g., \cite{ste91}; \cite{bun23}). 
If the origin of NUV radiation toward N4 is emission lines, C$_{\rm III}$] and Mg$_{\rm II}$ become candidates for the NUV radiation. 
\citet{son97} monitored the intensity of several line emissions including C$_{\rm III}$] emission at UV/NUV band after outburst in SN1987A. 
They found that the intensity of all of observed line emissions peaked at $\sim$400 days after the outburst, and disappeared at $\sim$2000 days.
This indicates that the radiation timescale of line emissions in the UN/NUV bands is several years after a strong shock. 
Present NUV radiation seen in N4 suggests that N4 is now under strong shock condition.  
If the age of W 50 is $\sim$10$^5$ years \citep{pan17}, the blast wave of the supernova that formed W 50 has already reached the boundary of the cavity created by the massive star before it becomes a compact star, and line emissions in the UV/NUV bands due to the effect of the supernova have already disappeared. 
This suggests that the supernova is not the origin of NUV radiation toward N4. 

Figure 4 shows a detailed comparison of the spatial distribution of NUV radiation with (a) integrated intensity of $^{13}$CO($J$=3--2) emission, (b) peak $T_{\rm MB}$ ratio of $^{12}$CO($J$=3--2)/$^{12}$CO($J$=1--0) emission, and (c) AKARI N160 \citep{doi15}.  
In Figure 4(a), the intensity of NUV decreases at the intense $^{13}$CO($J$=3--2) emitting area around ($\alpha$, $\delta$)(J2000) $\sim$ (\timeform{19h10m10.5s}, \timeform{5D4'42''}).   
The intensity of the extended $^{13}$CO($J$=3--2) emitting area also shows an anti-correlation with that of NUV radiation.  
In Figure 5, the peak $T_{\rm MB}$ ratio higher than 0.6 extends over N4, indicating that the CO molecules in N4 are relatively excited compared to the typical molecular clouds in the Galactic Plane (e.g., \cite{sa08b}). 
Particularly, the peak $T_{\rm MB}$ ratio in Figure 4(b) and Figure 5 is as high as $\sim$0.9 where the intensity of NUV radiation decreases.  
The region with a $T_{\rm k}$ of $\sim$56 K estimated in \citet{yam22} corresponds to the region where the intensity of NUV decreases. 
In Figure 4(c), far-infrared radiation peaks in the region where NUV radiation decreases at ($\alpha$, $\delta$)(J2000) $\sim$ (\timeform{19h10m10s}, \timeform{5D4'44''}). 
These results suggest that NUV radiation is absorbed by interstellar dust in N4 and the heated dust reradiates in the far-infrared wavelength.  
If this is true, NUV radiation is radiated from behind the molecular cloud. 
Molecular cloud is heated up by photoelectric heating through electrons released due to the photoelectric effect caused by the phenomenon where interstellar dust absorbs NUV radiation. 
As mentioned in section 4, while N3 is also situated within the radio continuum shell \citep{liu20}, no NUV radiation has been detected. 
$T_{\rm k}$ of N3 has been estimated to be $\sim$30 K, which is higher than that of typical molecular clouds in the Galactic plane.
They conclude that the heating source for N3 is cosmic-rays, not X-rays, as the heating rate from the X-ray radiation is three orders of magnitude lower than the cooling time scale.
In contrast, the heating source for N4 is identified as an X-ray jet.
This difference in heating sources may influence the physical phenomena that produce NUV radiation.
The details of this will be discussed in a forthcoming paper.

Figure 6 shows a schematic view of the current condition of N4. 
The results discussed in this section, along with the findings in \citet{yam22}, suggest that N4 is currently interacting directly with the jet from SS 433. 
Atoms of carbon and/or magnesium etc. are ionized and excited in the interacting layer between the jet and N4.
Then, those emissions are detected as the NUV radiation with GALEX.
Figure 7 presents a three-color composite image, consisting of red: H$\alpha$ emission by IPHAS \citep{bar14}, green: an intensity map of $^{12}$CO($J$=1--0) emission integrating from 45 to 60 km s$^{-1}$, and blue: a map identical to the green one but integrating from 25 to 35 km s$^{-1}$.
The molecular cloud depicted in blue is identified as a giant molecular filament situated in front of the W 50 /SS 433 system, with an estimated distance of $\sim$1.7 kpc \citep{lin20}. 
This foreground component overlaps with two-thirds of N4 at the southeastern part.
The $N$(H$_{\rm 2}$) of the foreground component is estimated to be a few times 10$^{21}$ cm$^{-2}$, corresponding to $A_{\rm v}$ of 2 to 5.
Considering $A_{\rm v}$ in both the foreground component and N4, as well as the number of stars identified in the H$\alpha$ image, both inside and outside N4, H$\alpha$ emission does not seem to be significantly dimmed by the molecular clouds.
Therefore, it is inferred that in N4, H$\alpha$ emission is not being radiated.
This may result by the moderate effective interacting velocity between the jet and N4 because alfven's critical ionization velocity of H$_{\rm II}$ and C$_{\rm III}$ are $\sim$50.9 km s$^{-1}$ and $\sim$19.7 km s$^{-1}$, respectively. 
This velocity range of the shock is similar to the velocity simulated in \citet{leh20}.
In the case of Cygnus loop \citep{dan01}, emissions of ionized atoms of oxygen, carbon, nitrogen etc., provide insights into the detailed shock conditions in the interacting layer between the cavity wall and the blast wave of the supernova.  
Thus, to gain a better understanding of the details of the interacting layer between the jet and N4, spectroscopic observations in the UV/NUV bands are essential.
The interacting layer of the jet and N4 is an important site of shock-cloud interaction. 
The region of shock-cloud interaction is the place where high-energy gamma-ray radiation is produced as a result of the collision between cosmic-ray protons accelerated by diffusive shock acceleration \citep{fer49} and the ISM.
Strong and long-term shock-cloud interaction continues while the jet emanates from SS 433, and the timescale of shock-cloud interaction in the present case is longer than that in the case of supernova remnants. 
This implies that the energy of the cosmic-ray protons may increase beyond 1 PeV, and N4 may be a candidate for a PeVatron.

\clearpage


\begin{figure*}[t]
  \begin{center}
    \includegraphics[width=160mm]{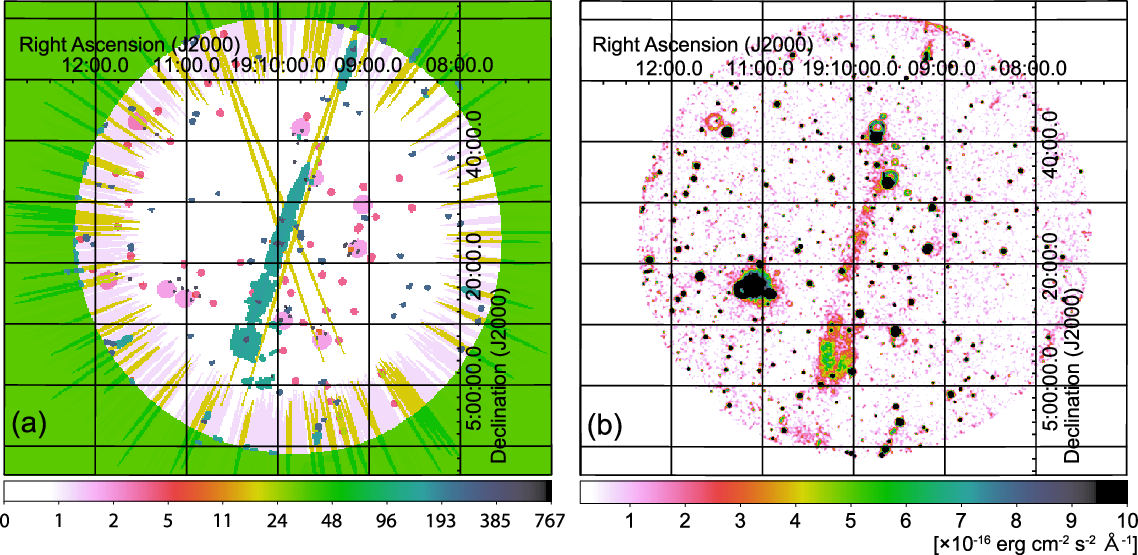}
  \end{center}
  \caption{(a) An artifact flag map in the field where present region is included. The number represents the summation of artifact effects occurred in ecah pixel shown in table 1.  (b) GALEX data image of the field. \label{fig:f1}}
\end{figure*}

\begin{figure*}[t]
  \begin{center}
    \includegraphics[width=160mm]{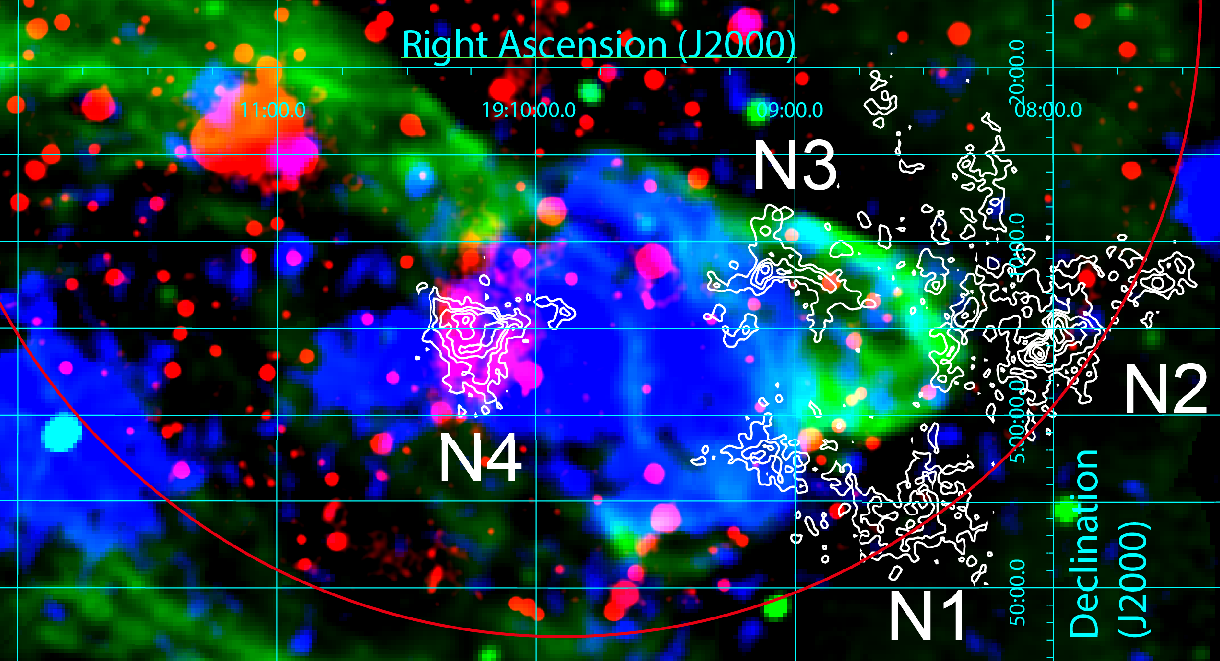}
  \end{center}
  \caption{Three composite color image of red: NUV radiation taken with GALEX, green: 1.4 GHz radio continuum radiation taken with VLA, and blue: 0.1--2.4 keV radiation taken with ROSAT superposed on integrated intensity map of $^{12}$CO($J$=1--0) emission taken with NRO45m telescope in contours. The contour levels of $^{12}$CO($J$=1--0) emission are illustrated from 9.26 K km s$^{-1}$ every 11.12 K km s$^{-1}$. A part of the red circle shows the edge of a field of view of GALEX observation. N1 to N4 are labels of molecular clouds defined in Yamamoto et al. (2008). \label{fig:f2}}
\end{figure*}

\begin{figure}
  \begin{center}
    \includegraphics[width=80mm]{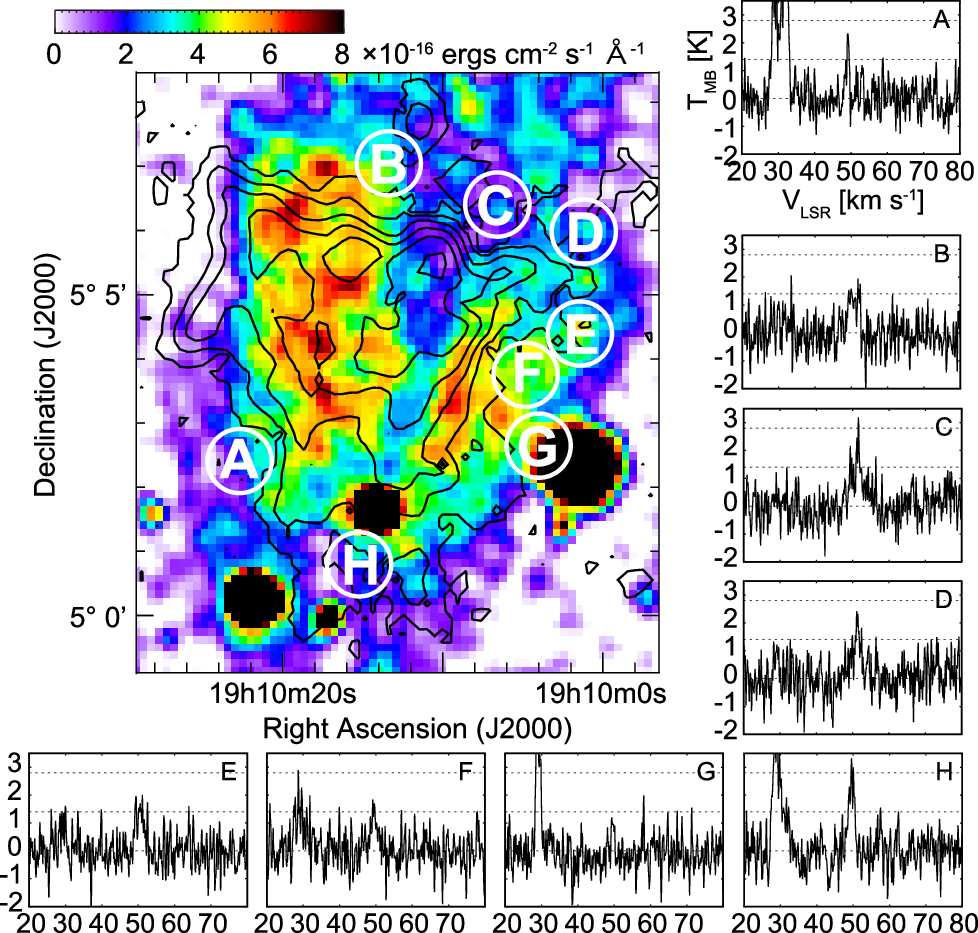}
  \end{center}
  \caption{Upper left: intensity distribution of NUV radiation in image superposed on peak $T_{\rm MB}$ distribution of $^{12}$CO($J$=1--0) emission in contours. The contour levels are illustrated from 2.8 K every 5.6 K. 
Circles labeled A to H indicate the positions of spectra in $^{12}$CO($J$=1--0) represented in the figure. 
The Spectra shown in the figure were generated by smoothing data within each circle, having a one-arcmin diameter, while the resolution of data used in this analysis is 20$\arcsec$. 
In each spectral panel, the lower horizontal dashed line represents the 0 K level, the middle horizontal dashed line signifies the r.s.m. levels of the datasets before smoothing data within each circle, and the upper horizontal dashed line corresponds to the lowest contour level of the peak $T_{\rm MB}$ distribution in the upper-left of the figure.  \label{fig:f3}}
\end{figure}

\begin{figure*}[t]
  \begin{center}
    \includegraphics[width=160mm]{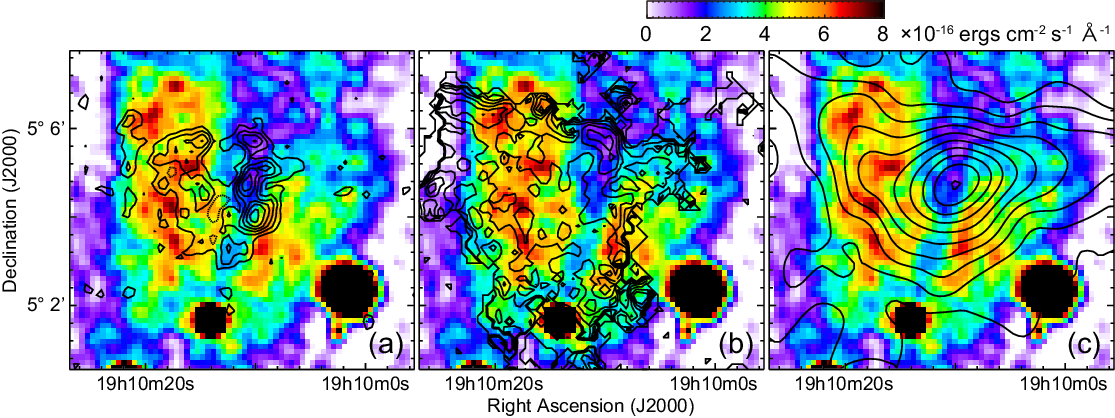}
  \end{center}
  \caption{Comparison of NUV with (a) integrated intensity of $^{13}$CO($J$=3--2) emission, (b) peak $T_{\rm MB}$ ratio of $^{12}$CO($J$=3--2)/$^{12}$CO($J$=1--0) emission, and (c) Akari N160 (160$\mu$m). Contour levels are (a) from 2 K km s$^{-1}$ every 1.5 K km s$^{-1}$, (b) from 0.3 every 0.1, and (c) from 330 MJy sr$^{-1}$ every 30 MJy sr$^{-1}$, respectively. The dotted contours in (a) represent the intensity decreases from the outside to the inside.  \label{fig:f4}}
\end{figure*}

\begin{figure}
  \begin{center}
    \includegraphics[width=80mm]{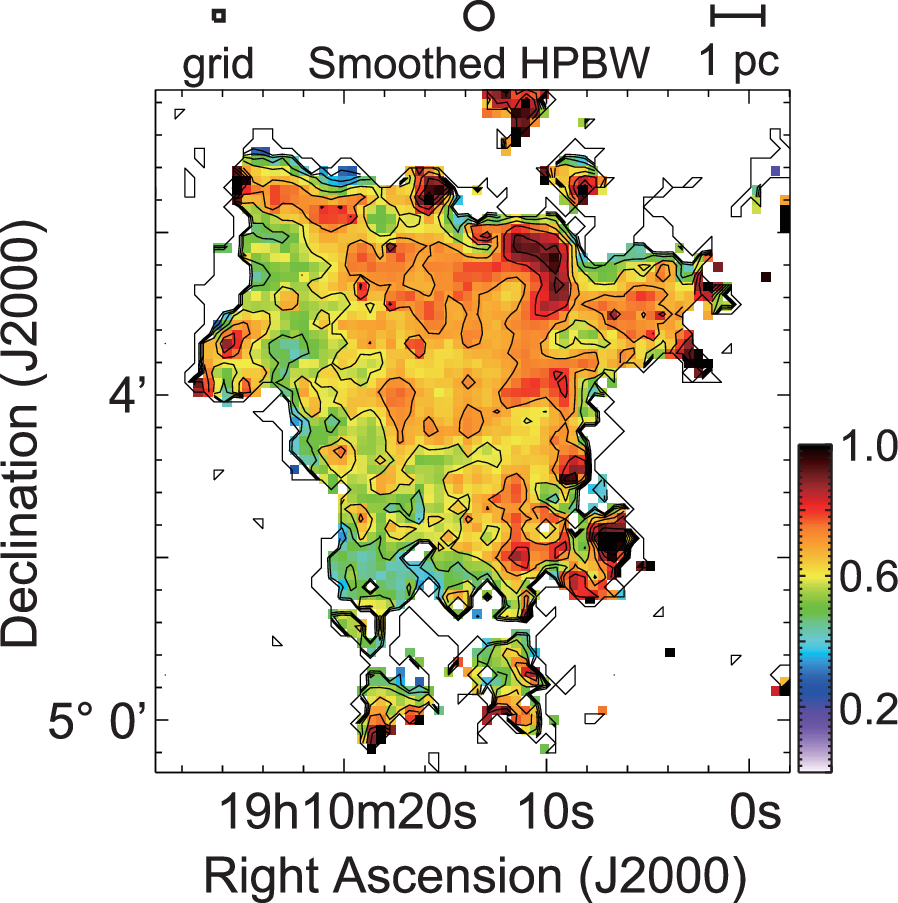}
  \end{center}
  \caption{Peak $T_{\rm MB}$ ratio of $^{12}$CO($J$=3--2)/$^{12}$CO($J$=1--0) emission. Contour levels are from 0.3 every 0.1. \label{fig:f5}}
\end{figure}

\begin{figure}
  \begin{center}
    \includegraphics[width=80mm]{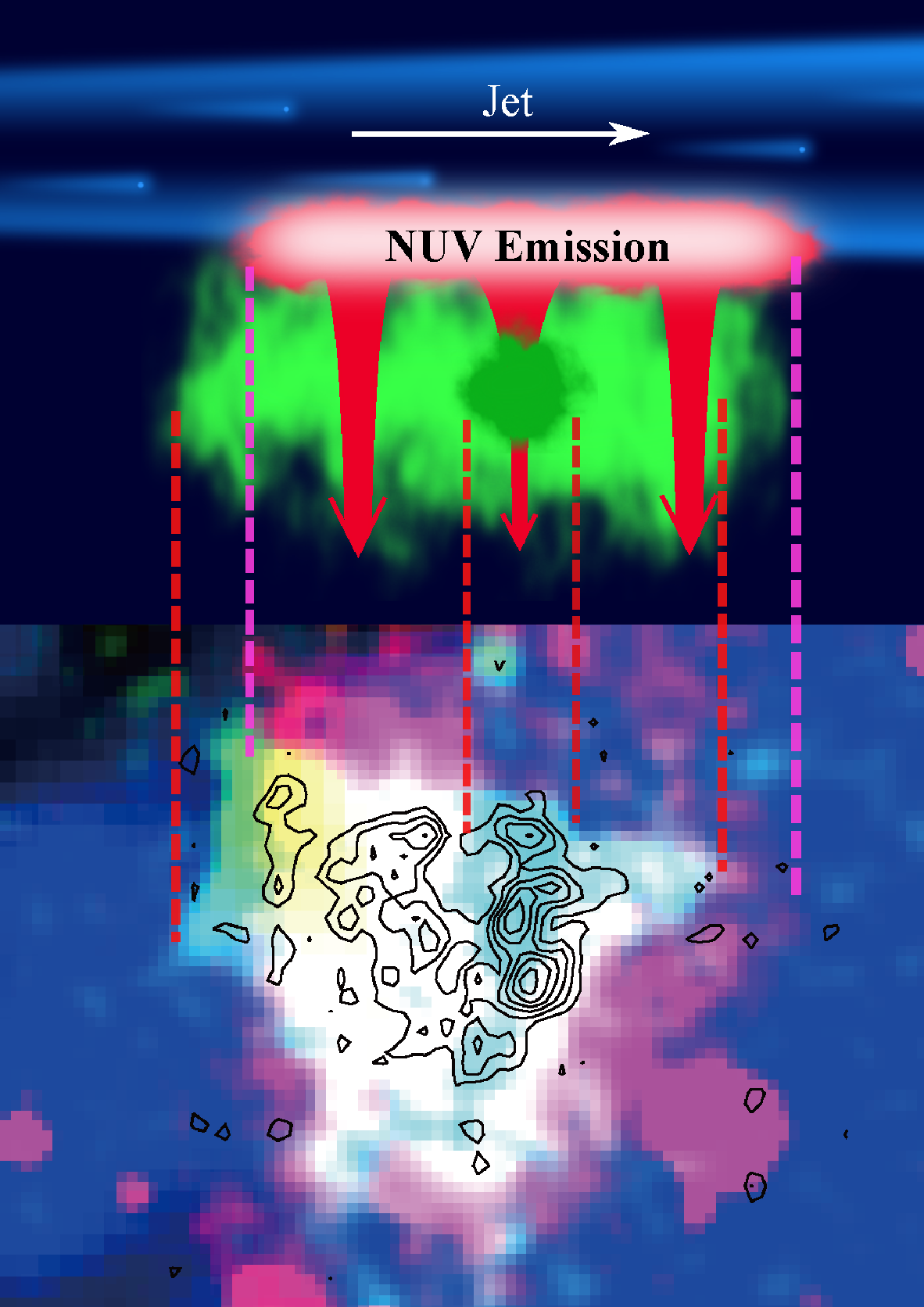}
  \end{center}
  \caption{Upper: Schematic view of nature of the interaction of the jet from SS 433 and N4. Colors of red, green, and blue represent NUV radiation,  molecular cloud traced in $^{12}$CO($J$=1--0), and the jet from SS 433, NUV radiation, respectively. Observers are at downward in the figure. Lower: Three composite color map of observational datasets. What colors of red, green, and blue indicate is the same as colors in upper figure. Contours represent integrated intensity of $^{13}$CO($J$=3--2) emission. Contour levels are the same as those in figure 3(a). Horizontal dashed lines represent edge of radiating area of NUV, molecular cloud, and the region where intensity of NUV decreases compared to surroundings in the figure.  \label{fig:f6}}
\end{figure}

\begin{figure}
  \begin{center}
    \includegraphics[width=80mm]{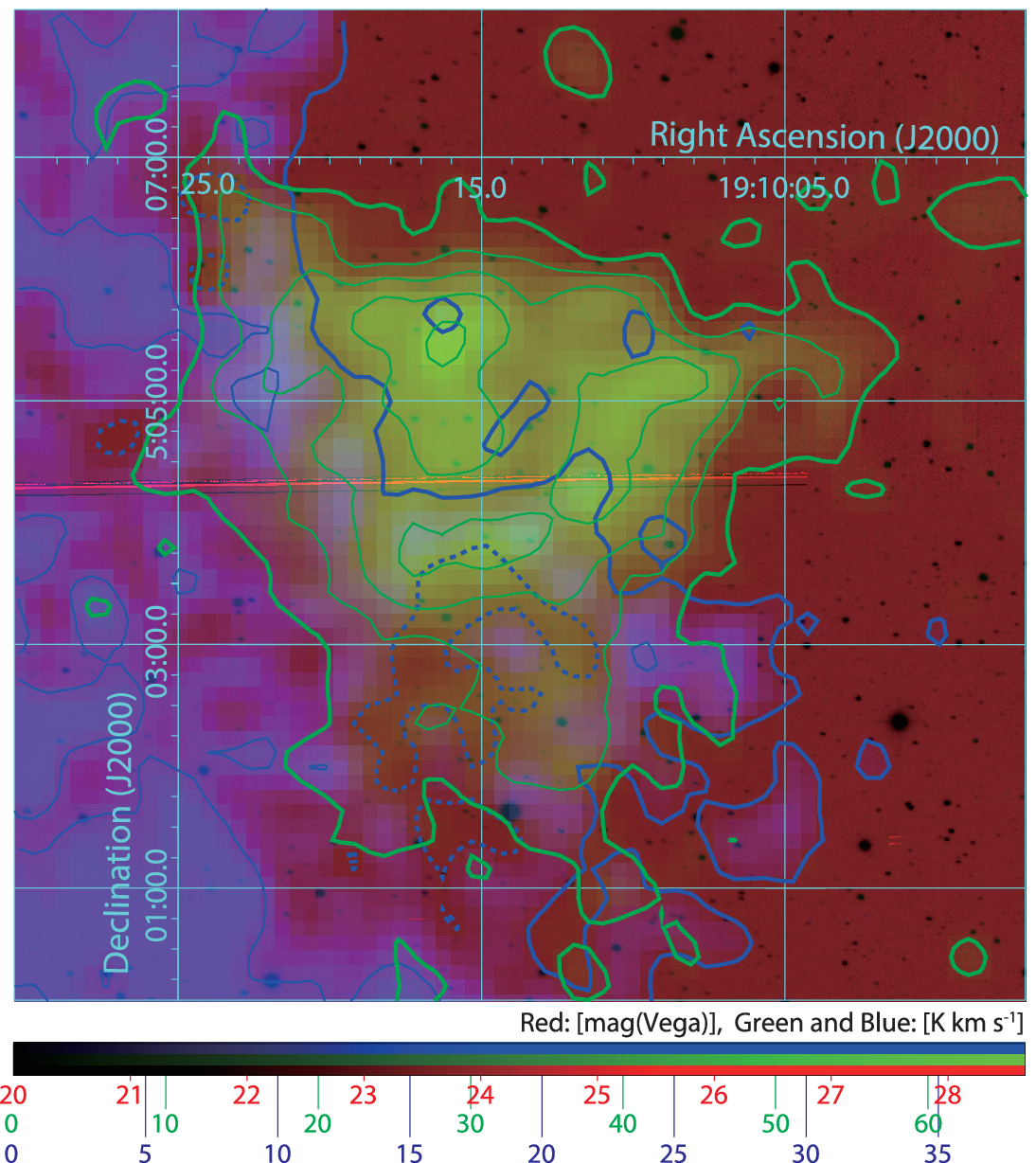}
  \end{center}
  \caption{Three composite color image of red: H$\alpha$ image by IPHAS \citep{bar14}, Geen: an intensity of $^{12}$CO($J$=1--0) emission integrating 45 to 60 km s$^{-1}$, and blue: same as green but integrating 25 to 35 km s$^{-1}$. CO intensities are also presented in Contours. Contour levels are the same as those in figure 1. The lowest level of the contours are shown by thick ones and othres are shown by thin ones. The dotted contours in blue shows that intensities decrease from outside to inside. \label{fig:f7}}
\end{figure}

\clearpage


\begin{table}
  \caption{Details of Artifact Flags of Galex data}
  \begin{tabular}{rl}
  \hline
  \hline
  Number & description \\
  \hline
  1 & (edge) Detector bevel edge reflection. \\
  2 & (window) Detector window reflection. \\
  4 & (dichroic) Dichroic reflection. \\
  8 & (varpix) Variable pixel based on time slices. \\
  16 & (brtedge) Bright star near field edge.  \\
  32 & Detector rim(annulus) proximity ( \>0.6 deg fld ctr).  \\
  64 & (dimask) dichroic reflection artifact mask flag.  \\
  128 & (varmask) Masked pixel determined by varpix.  \\
  256 & (hotmask) Detector hot spots.  \\
  512 & (yaghost) Possible ghost image from YA slope. \\
  \hline
  \end{tabular}
\end{table}


\begin{ack}
This work has been supported by the Japan Society for the Promotion of Science (JSPS) Grants-in-Aid for Scientific Research (19K03912, 21H01128). 
This work has also been supported in part by the Collaboration Funding of the Institute of Statistical Mathematics ``New Perspective of the Cosmology Pioneered by the Fusion of Data Science and Physics''. 
This research is based on observations made with the GALEX mission, obtained from the MAST data archive at the Space Telescope Science Institute, which is operated by the Association of Universities for Research in Astronomy, Inc., under NASA contract NAS 526555.
The James Clerk Maxwell Telescope is operated by the East Asian Observatory on behalf of The National Astronomical Observatory of Japan; Academia Sinica Institute of Astronomy and Astrophysics; the Korea Astronomy and Space Science Institute; Center for Astronomical Mega-Science (as well as the National Key R\&D Program of China with No. 2017YFA0402700). Additional funding support is provided by the Science and Technology Facilities Council of the United Kingdom and participating universities and organizations in the United Kingdom and Canada.
The Nobeyama 45 m radio telescope is operated by Nobeyama Radio Observatory, a branch of National Astronomical Observatory of Japan. 
This research is based on observations with AKARI, a JAXA project with the participation of ESA.
This paper makes use of data obtained as part of the INT Photometric H$\alpha$ Survey of the Northern Galactic Plane (IPHAS, www.iphas.org) carried out at the Isaac Newton Telescope (INT). 
The INT is operated on the island of La Palma by the Isaac Newton Group in the Spanish Observatorio del Roque de los Muchachos of the Instituto de Astrofisica de Canarias. 
All IPHAS data are processed by the Cambridge Astronomical Survey Unit, at the Institute of Astronomy in Cambridge. 
The bandmerged DR2 catalogue was assembled at the Centre for Astrophysics Research, University of Hertfordshire, supported by STFC grant ST/J001333/1.
\end{ack}

\end{document}